\documentclass[letterpaper, 11pt]{article}
\usepackage{amsmath, amssymb}
\usepackage{bm}
\usepackage{graphicx}
\usepackage[font={small,it}, labelfont=bf]{caption}                
\usepackage[margin=1in]{geometry}                         
\usepackage{setspace}                               
\usepackage{wrapfig}                                
\usepackage{xcolor}
\usepackage{multirow,tabularx}
\newcolumntype{Y}{>{\centering\arraybackslash}X}

\usepackage{cite}

\usepackage{titlesec} \titleformat*{\section}{\bfseries}              
\titleformat*{\subsection}{\bfseries}                       

\usepackage{authblk}
\author[1]{Soojung Baek}
\author[1]{Kristofer G. Reyes\thanks{kreyes3@buffalo.edu}}
\affil[1]{Department of Materials Design and Innovation, University at Buffalo}

\title{Problem-Fluent Models for Complex Decision-Making in Autonomous Materials Research}
\date{}

\begin{document}
\maketitle

\begin{abstract}
    We review our recent work in the area of autonomous materials research,
    highlighting the coupling of machine learning methods and models and more
    problem-aware modeling. We review the general Bayesian framework for
    closed-loop design employed by many autonomous materials platforms. We then
    provide examples of our work on such platforms. We finally review our
    approaches to extend current statistical and ML models to better reflect
    problem specific structure including the use of physics-based models and
    incorporation operational considerations into the decision-making procedure.
\end{abstract}

\section{Introduction}
\label{sec:introduction}

\begin{figure}
\centering
    \includegraphics[width=0.8\textwidth]{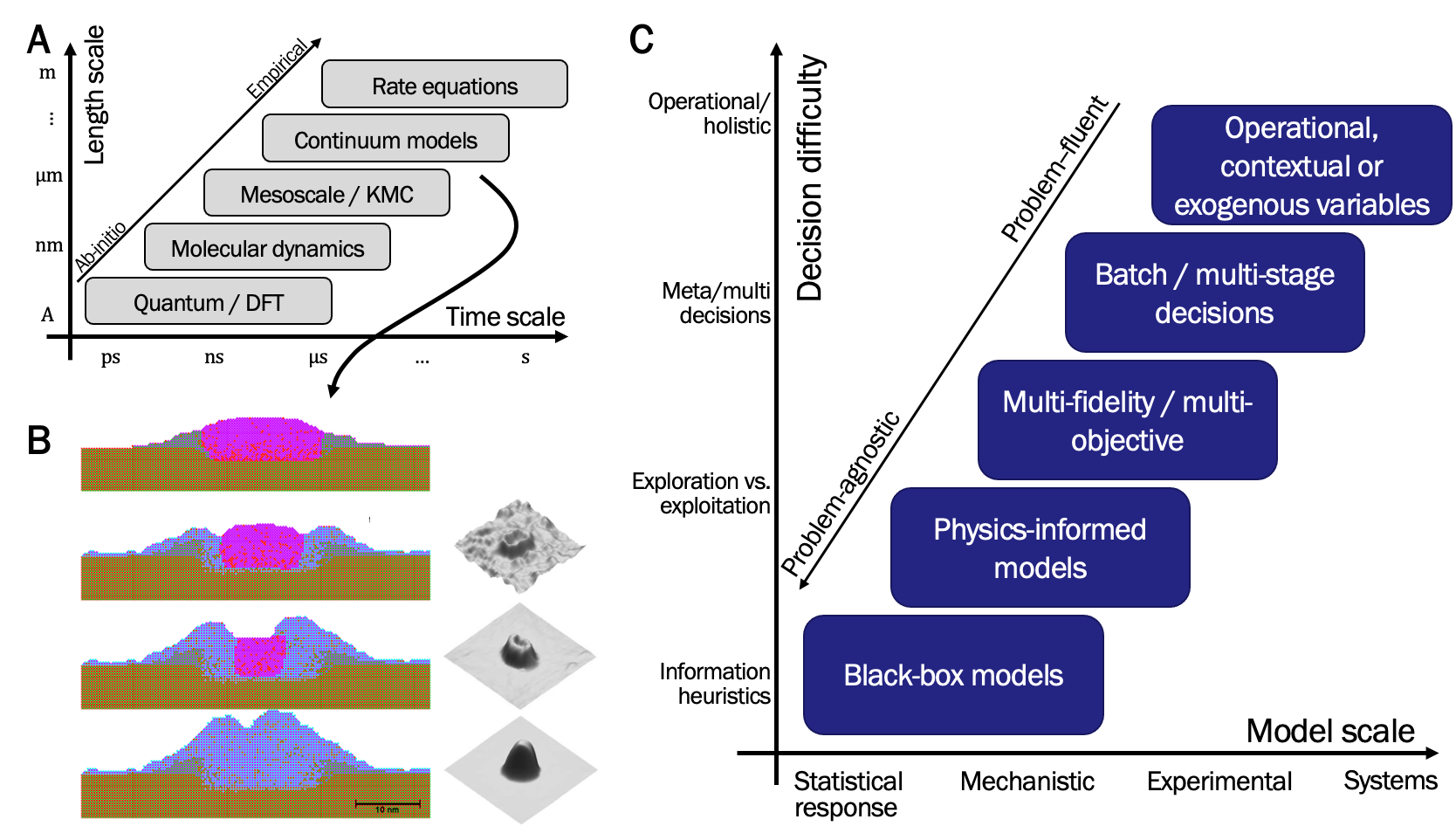}
\caption{A) Scale comparison of methods from traditional computational
materials. B) Example of multi-phase kinetic Monte Carlo simulations of quantum
dot formation (taken from \cite{reyes2013unified}).  C) An analogous scale
comparison for autonomous materials research systems.} 
\label{fig:scales}
\end{figure}

Practitioners of computational materials science know well the figure in which
computational techniques and methodologies are plotted against the time and
length scales for which they are typically deemed to be appropriate. This
figure, reproduced in figure \ref{fig:scales}A is known by practitioners to
provide a qualitative understanding of the overlapping domains of the
respective methodologies, implying a continuum of models that offer various
strengths and weaknesses for addressing specify types of materials modeling
problems. While open to some interpretation, figures such as these are meant to
illustrate a broad range of phenomena that can be captured, with each method
achieving its own balance between \emph{ab initio} calculation and
empirical modeling.

Indeed, our early work modeling of more complex phenomena within lattice-based
kinetic Monte Carlo (KMC) atomistic simulations (figure \ref{fig:scales}B) is
one example of attaining such a balance through empirically designed liquid
local neighborhoods calibrated alongside analytical continuum models. This
balance allowed us to simulate mesoscale, multi-phase phenomena, capturing the
evolution of nano- to micrometer features of materials synthesized over a
simulated period of$10^{-1}$ to $10^1$ seconds. With such models, we
were able to simulate nanoring formation and predict core-shell quantum dot
structures in nanomaterials grown via Molecular Beam Epitaxy
\cite{reyes2013unified, dejarld2013mechanisms, bollani2014ordered}, study
kinking and faceting phenomenon in nanowire Vapor-Liquid-Solid (VLS) and
Vapor-Solid-Solid (VSS) synthesis, and study the evolution of porous material
under high-temperature annealing \cite{reyes2014fast}.  Through the proper
modeling of relevant problem-specific features encompassing a spectrum of
atomistic and continuum treatments, we could better capture physical phenomena
leading to more insightful simulations at the scales of interest.

Recently, materials science has seen an influx of machine-learning (ML) methods
to gain similar insight using data-driven statistical models rather than
physical ones. Such methods allow us to study and predict material structure
and properties, analyze complex and rich data from materials experiments and
even perform autonomous, closed-loop research to design and discover optimal or
new materials. Due to the computational nature of such ML techniques, which
often require high-performance computing to build, train, and use such models,
their application to materials science problems should qualify as a nascent
budding branch of computational materials science.

Part of the attractiveness of many ML algorithms is their general-purpose
nature. Given a large enough data set, it often suffices to treat any
ML-enabled materials science task as the curation of yet-another-data set to
train and subsequently apply problem-agnostic models and methods to make
predictions, perform materials screening, or analyze data. Yet this
problem-agnostic perspective fails to capture the richness of the structure of
many materials science tasks. For example, the data requirements of many ML
methods, such as deep learning models that often require tens-of-thousands to
millions of data points, may be prohibitively too large in a setting where such
data must be obtained through laborious, error-prone, and time-intensive
physical or computational experiments. As another example, much of the data may
be explained through mechanistic physical models that, if utilized correctly,
could offset such ostensibly large data requirements. Indeed, such a
problem-agnostic, purely data-centric perspective of the application of ML to
materials science is analogous to empirical phenomenological models in
traditional computational materials science.  Extending this analogy, we argue
that, like the continuum between \emph{ab initio} and empirical methods of
traditional computational materials, the use of ML models and methods have a
similar balance between generic, problem agnosticism and more problem-fluent
models that are aware of the many subtleties of the materials problem at hand.

This is particularly important in the area of autonomous materials research.
This field has garnered recent attention due to the demonstration of
several autonomous robotic materials development platforms
\cite{nikolaev2016autonomy, chang2020efficient, kusne2020fly,
epps2020artificial, gongora2020bayesian, roch2020chemos}. In such systems, the
robot iteratively and efficiently builds knowledge of a material system through
strategically selected experiments, the results of which inform the selection
of subsequent experiments. While many of these initial systems demonstrate
proofs-of-concept using generic, off-the-shelf models and decision-making
strategies, this field is primed to further explore problem-aware modeling. The
hope in developing such modeling is to increase the efficiency of autonomous
research campaigns and provide the autonomous system more agency in making many
higher-order decisions, which are often ignored in the proof-of-concept systems
currently in production. 

The development of such models and decision-making techniques is
depicted in figure \ref{fig:scales}C. The figure illustrate the breadth of
problem-fluency as a function of an autonomous platform's models of a material
system (including what constitutes the ``system" itself), along with the
considerations it makes when deciding experiments to run.  In the lower-left
corner, such platforms use empirical, black-box models of experimental
responses and make decisions through simple heuristics such as randomized or
space-filling designs. At the next level, models incorporate more sophisticated
features such as uncertainty quantification and the use of, or inference on
physics-based knowledge. With such models, autonomous agents may make
correspondingly more complex decisions, including balancing between a)
resolving uncertainties of the models or inferring a better understanding of
physical mechanisms underpinning observed responses and b) selecting
experiments that achieve experimental objectives based on best estimates --
this is the \emph{Exploration vs. Exploitation Dilemma} that we outline below.

At the next level, models include multiple aspects of running an
experiment such as a) joint models for multi-modal materials characterizations
that may be performed on a sample, b) the ability to run physics simulations,
and c) a quantification of the fidelities from various measurements that can be
made, either \emph{in silico} or physically.  This more fluent understanding
grants an autonomous platform further agency when making decisions.  For
example various ``multi-" methods such as multi-objective optimization
\cite{miettinen2012nonlinear, zitzler1999multiobjective, shimoyama2013kriging},
multi-fidelity models \cite{peherstorfer2018survey}, and multi-information
source \cite{poloczek2017multi, lam2015multifidelity} techniques allow agents
to not only choose experimental inputs such as synthesis parameters, but also
modes of characterization or evaluation, such as choosing to run an experiment
versus a simulation and, if tunable, at which fidelity level to run such
simulations.

Finally, at the last tier models include  more operational or
systems-level aspects of running an entire experimental campaign. For example,
we can consider modeling a) experimental costs, b) personnel, material or
equipment availability, or c) the ability to run campaigns within a
heterogeneous network of manual, semi-autonomous and fully autonomous materials
experimental apparatuses.  With a high-level perspective of experimental
campaigns, the decisions that an agent may make here encompass several scales,
from deciding individual synthesis conditions that affect a material at the
nanoscale, to broad resource allocation decisions that impact the efficiency or
success of an entire research campaign.

Through this delineation, we wish to make explicit the analogy 
with the continuum of models and algorithms from traditional computational
materials.  As in the choice between empirical and \emph{ab initio} methods,
the choice between black-box methods to more problem-fluent ones comes with a
trade-off in model simplicity and ease of computation.  However, unlike more
traditional computational materials, this continuum of autonomous materials
techniques as not been explored much.  Research in this exploration could
result in better modeling and decision-making, allowing us to build more
holistic autonomous materials platforms that can achieve accelerated materials
design and discovery.

In this article, we review some of our recent progress in this
direction. In section \ref{sec:review}, we first review the generic framework
work we use to model closed-loop autonomous materials research using Bayesian
statistics and decision-making policies. We then highlight in section
\ref{sec:examples} a few examples of our collaborations developing robotic
autonomous materials research platforms to illustrate what is possible using
only generic, off-the-shelf models and methods. We then introduce examples of
integrating problem-specific structure and fluency in both the modeling and
decision-making employed by such autonomous research systems. These examples
are described in detail in section \ref{sec:complex}, and include a) the
inclusion of physics-based models in concert with ML models and
decision-polices, b) the incorporation of more complex decision-making
structure to better capture typical decisions made in materials science
research, and c) the introduction of operational considerations such as time
and costs often encountered when running physical experimental campaigns.

\section{Closed-loop design and Autonomous Materials Development}
\label{sec:review}

Here we review the basic framework we use for closed-loop autonomous materials
development. We use a combination of Bayesian statistical models to represent
beliefs about the material systems and use algorithms that select one or more
experiments to run based on those beliefs. Experimental observations or
responses are then used to update the models, closing the loop. This loop is
repeated several times throughout an experimental campaign until some
termination criteria are satisfied. Throughout, we usually speak of running an
experiment or performing some experimental action. Typically, this means
performing an actual physical experiment, though this could also mean running a
computer simulation or evaluating some surrogate model of the experiment. 

What we present here, which we will call generically ``closed-loop
design", has significant overlaps with many models and techniques, which we
review in broad strokes. The field of Optimal Experimental Design (OED)
considers the selection of experiments with optimal information content,
through various definitions of optimality \cite{fisher1936design,
kennard1969computer}. Active Learning (AL) \cite{settles2009active,
cohn1994improving, cohn1996active}, Optimal Learning (OL)
\cite{ryzhov2012knowledge, powell2012optimal, powell2008optimal}, and Bayesian
Optimization (BO) \cite{snoek2012practical, frazier2018tutorial} frame the
question of informationally optimal experiments within an iterative loop, with
each technique using different models representing the system under study and
different decision-making policies geared for various experimental objectives.
For example, AL is often concerned with reducing uncertainty, while BO and OL
policies are geared toward optimizing experimental responses.  BO is also used
as a global optimization algorithm
\cite{klein2017fast,springenberg2016bayesian, mockus2012bayesian}, occupying a
similar niche as methods like genetic algorithms \cite{hassan2005comparison,
wright1991genetic}. Optimal Control (OC) \cite{bertsekas1995dynamic,
zhou1996robust} considers the problem of driving a system to a specific state,
usually under the assumption of known dynamics.  Reinforcement Learning (RL)
\cite{sutton2018reinforcement, kaelbling1996reinforcement} considers a similar
state-dependent control problem, but where some aspect of the dynamics or the
effect on the dynamics upon taking some action must be learned through
optimally selected actions and closed-loop feedback is obtained through
incremental rewards. Such techniques overlap in their modeling of the system
being queried, the design of actions and experiments to work toward some
objective, and the ability to receive feedback from such actions.

\begin{figure}
\centering
\includegraphics[width=0.8\textwidth]{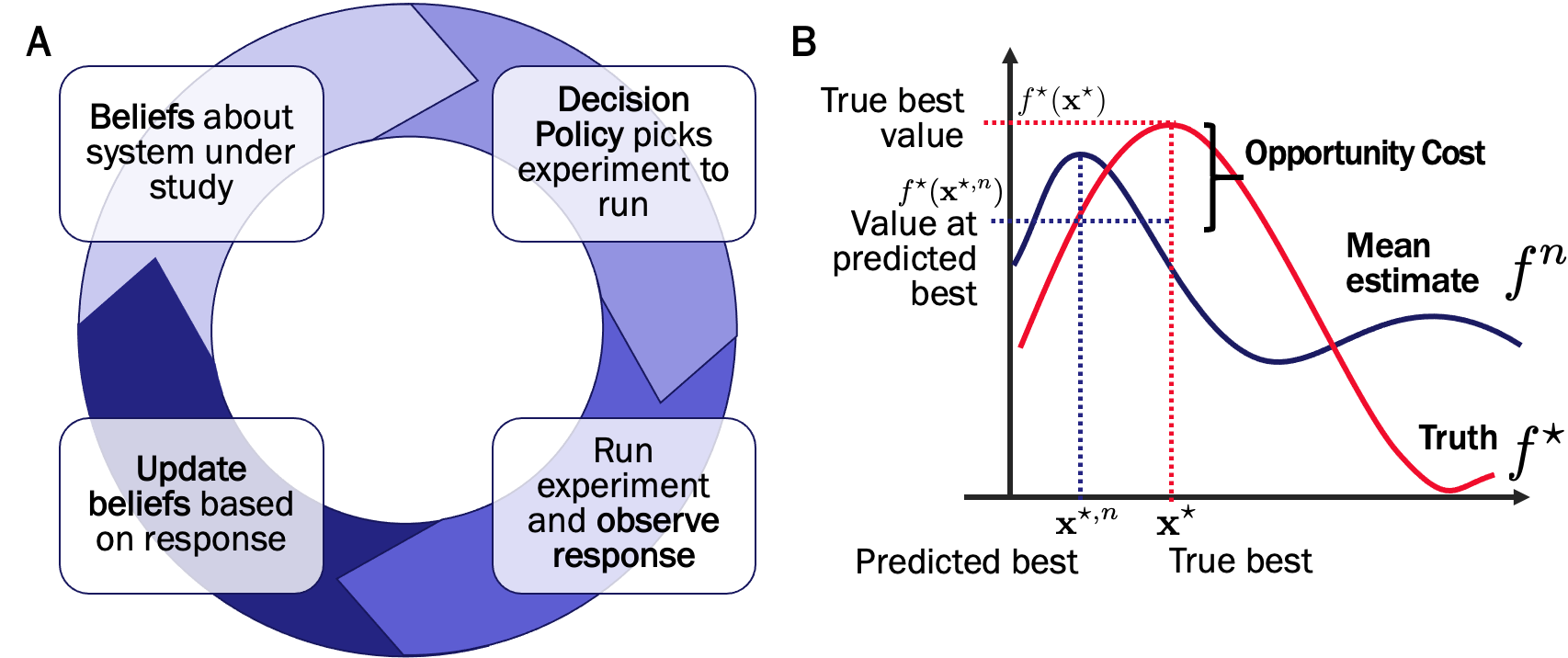}
\caption{A) A schematic of the closed-loop design main loop employed by several
    autonomous materials research platforms. B) An illustration of a key metric
    known as Opportunity Cost used to assess the performance of decision-making
    policies and models.}
\label{fig:closed_loop_design}
\end{figure}

\subsection{Basic Formalization}
Our approach to closed-loop materials design and autonomous materials
development consists of a few basic ingredients, illustrated in figure
\ref{fig:closed_loop_design}A. We first assume some parameterization of an
action or experiment, \(\mathbf x\). Upon taking action (such as running an
experiment), we assume an experimental response or observation of the form
\begin{equation}
\hat f(\mathbf x) = f^\star(\mathbf x) + W.
\label{eqn:noise}
\end{equation}
Here we have written the observed response \(\hat f(\mathbf x)\) is a
random, noisy perturbation of the ground-truth response function
\(f^\star(\mathbf x)\). Sometimes, we assume some notion of the state \(S\) of
the experimental campaign, and may sometimes write the ground-truth response as
additionally a function of the state: \(f^\star(\mathbf x) = f^\star(\mathbf x;
S)\). We also model some unknown quantities \(\bm{\theta}^\star\) of the system
being studied.  This unknown could directly be the ground truth response itself
\(\boldsymbol \theta^\star = f^\star\) but can represent other unknowns of the
system. For example, we could presume that the response may be mechanistically
or empirically captured using some parameterized forward model
\(f^\star(\mathbf x) = f_M(\mathbf x; \bm{\theta}^\star)\), in which the
unknown quantities represent the effective true value of such parameters.

We assume that if we knew the unknown quantities \(\bm{\theta}^\star\)
perfectly, we would not have to run any additional experiments. That is,
perfect knowledge of these quantities would be all that is needed to
compute experimental responses \emph{in silico}, to a high degree of
fidelity. However, estimates of \(\bm{\theta}^\star\) must be derived
from a finite and often limited set of imperfect experimental
observations. As such, we never truly have perfect knowledge of the
unknown quantities, even having completed several experiments. Instead,
knowledge of these unknowns will be modeled to capture such uncertainty.
For this, we rely on a Bayesian interpretation: we view the unknowns
\(\bm{\theta}^\star\) probabilistically, as Random Variables (RVs).
Rather than provide a point estimate of what we believe the unknowns to
be at any particular stage of an experimental campaign, we instead
assign the RVs an entire probability distribution, which captures our
uncertain beliefs about the true values of \(\bm{\theta}^\star.\) Upon
receiving additional data, we may update these beliefs to reflect a
combination of this new data along with what was believed before this
data was observed. This combination results in a posterior belief that
blends priors with data in a rigorous way \cite{gelman2013bayesian}.
This Bayesian update is what ``closes'' the loop in closed-loop
autonomous materials research. Depending on the distribution selected to
represent the beliefs on \(\bm{\theta}^\star\) as well as what is
directly observed, performing this Bayesian update could be as easy as
evaluating some closed-form formula or may require the use of
computationally intensive Monte Carlo methods such as Markov Chain Monte
Carlo \cite{green1995reversible, gamerman2006markov}.

The random variable interpretation for \(\bm{\theta}^\star\) captures a
variety of perspectives for what is deemed to be unknown. If
\(\bm{\theta}^\star\) represents a finite set of quantities, such as
unknown effective parameters to a model, then Bayesian beliefs can be
captured through multivariate probability distributions. If
\(\bm{\theta}^\star\) represents some unknown function, such as in the
case where we directly represent beliefs of the ground-truth response
function \(\bm{\theta}^\star = f^\star\), then Bayesian beliefs are
probability distributions on random functions, i.e.~stochastic
processes. For example, perhaps one of the most popular ways of directly
representing some experimental response \(f^\star\) as a function of
experimental inputs is with Gaussian Process (GP) beliefs
\cite{williams1996gaussian, williams2006gaussian}, which we denote as

\[f^\star \sim \mathcal {GP}(\mu(\mathbf x), \Sigma(\mathbf x, \mathbf x^\prime)).\]

This is the functional or infinite-dimensional analog of a multivariate
normal (MVN) distribution, and like the MVN distribution, it is
parameterized by a mean or expected estimate \(\mu(\textbf x)\) of the
unknown function \(f^\star(\mathbf x)\), along with covariance
information captured by the bivariate function
\(\Sigma(\textbf x, \textbf x^\prime)\). The covariance function
contains information that qualifies the estimate provided by
\(\mu(\textbf x)\). For example, the diagonal terms
\(\sigma^2(\mathbf x, \mathbf x) = \Sigma(\textbf x, \textbf x)\)
specify the uncertainty associated to the estimate of the unknown
function \(\mu(\mathbf x)\) for any particular input \(\mathbf x\),
while off-diagonal terms \(\Sigma(\mathbf x, \mathbf x^\prime)\)
describe the assumed statistical relationship between the unknown
function values \(f^\star(\mathbf x)\) and
\(f^\star(\mathbf x^\prime)\).

Decision-making policies select the next action to take, given beliefs
about the unknown quantities \(\bm{\theta}^\star\). These policies
typically consider several factors when making such a decision,
including --- among other considerations --- the overall goal or
objective of running the closed-loop campaign, what we know about
\(\bm{\theta}^\star\), and the inherent variability of any response we
would observe upon taking the decided-upon action. Due to uncertainties
in our beliefs, most policies acknowledge the imperfect and uncertain
context in which they are asked to make decisions and often balance
between generally resolving uncertainties (called exploration) and
focusing on achieving the experimental objectives (called exploitation).
Achieving a harmonious balance between exploration versus exploitation
leads to an efficient, strategic, and intelligent exploration of
experiment space. The unknown quantities are learned to the minimal
level of fidelity needed to achieve the goals of a campaign.

Two typical objectives are 1) \emph{response optimization} and 2)
\emph{global learning}. In response optimization, the task is to
identify the input \(\mathbf x^\star\) whose resulting response is, on
average, optimal over the entire space of feasible inputs
\(\mathcal X\). In symbols, we wish to learn

\[x^\star = \max_{x \in \mathcal X} f^\star(x),\]

In global learning, the objective learns the unknown quantities
outright, i.e., learn the most accurate estimate of
\(\boldsymbol \theta^\star\). This could mean, as examples, learning the
response surface if \(\boldsymbol \theta^\star = f^\star\) or learning
the effective parameters to a physics-based model. Specific policies are
meant for specific experimental objectives. For example, in response
optimization, two policies we often use in our own work are the
Knowledge Gradient (KG) policy
\cite{frazier2008knowledge, frazier2009knowledge} and Expected
Improvement (EI) policy
\cite{ryzhov2016convergence, chick2010sequential}, both of which are
examples of Look-Ahead policies. Such policies attempt to estimate the
expected impact of running a potential experiment on future, posterior
beliefs, using current beliefs to calculate this expectation. Policies
for global learning include many of the so-called Alphabet-optimality
designs \cite{chaloner1995bayesian} and many active learning strategies.

While we have described the basic and generic framework for closed-loop
design, we shall see that the implementation of the different components
varies significantly in model and algorithm complexity and the ability
to capture the specific structure of a problem. We will highlight a few
examples of this in section \ref{sec:complex}. However, due to the
ubiquity of the generic problem of iterative decision-making under
uncertainty, many off-the-shelf techniques have been developed that can
be applied directly to autonomous materials research platforms to
perform, to first order, basic decision-making to optimize material
properties. In the next section, we will review some of our
collaborative work in developing such autonomous systems.

\section{Examples of closed-loop autonomous materials platforms}
\label{sec:examples}

This generic framework encompasses many closed-loop or active learning
scenarios in which Bayesian beliefs are interactively improved, and
experimental objectives are pursued in tandem. One of the most popular
forms of such closed-looped methods is black-box Bayesian optimization
(BO). In black-box BO, beliefs on the response function are typically
directly represented by GPs or other types of statistical or machine
learning models, and policies such as EI are used to select
information-rich experiments to identify the experimental inputs that
optimize the observed response in as few iterations around the
closed-loop as possible. Much of our work in BO has been implementing
and adapting such methods within the context of autonomous materials
systems.

For example, with Keith Brown at Boston University, we have applied this
black-box BO framework in optimizing mechanical properties of additively
manufactured mechanical structures using a Bayesian autonomous
researcher (BEAR) that consisted of an array of 3D printers, a robot
arm, and mechanical testing equipment \cite{gongora2020bayesian}. Using
GP beliefs models of structure toughness in tandem with the EI policy,
we had shown how the autonomous system efficiently explores a
four-dimensional parameterized structure space to identify those
structures that optimized certain mechanical properties, resulting in an
almost 60 times reduction in the number of structures tested when
compared to a traditional factorial design of experiments.

In another collaboration, with Milad Abolhasani at North Carolina State
University, we studied the use of an ensemble of models to represent
beliefs on experimental response functions to optimize properties of
colloidal quantum dots (QDs) synthesized by an autonomous flow reactor,
the Artificial Chemist \cite{epps2020artificial}. Rather than capture a
continuous space of response functions using a distribution on
functions, as with GPs, ensemble beliefs use a finite sample of
potential candidate response functions. Summary statistics, such as
expected values or covariances, typically calculated using a
distribution like a GPs, can be approximated using sample statistics
over the ensemble. Ensemble beliefs also allow for the use of more
complex models, which in the Artificial Chemist case consisted of
ensembles of neural networks. Using the autonomous Artificial Chemist
and a variety of BO-oriented policies, we studied the effectiveness of
this model and policies to quickly tune properties of flow-reactor
synthesized colloidal QD.

These and other examples point to the opportunities that even generic,
off-the-shelf models and algorithms such as GPs and EI can help
accelerate materials research through an intelligent and strategic exploration
of design space. Techniques and models such as GP beliefs and the EI
decision-making policy for black-box BO are generic. The same models and
methods have found popular applications elsewhere, apart from autonomous
materials research. For example, BO has experienced recent popularity as
a global-optimization method used to tune hyper-parameters of ML models
\cite{klein2017fast, snoek2012practical, wu2020practical} In this
application, hyper-parameter space is iteratively explored, and
different hyper-parameters are evaluated through an objective function
such as an error on validation or hold-out data sets. Finding a global
optimum of such an objective function could result in optimally
calibrated and validated models.

However, this example betrays one reason why we may wish to consider
models and methods beyond black-box BO for autonomous materials design.
In contrast to evaluating validation errors for some ML model, which is
all done \emph{in silico}, using generic BO methods in materials
research ignores many realistic aspects of running \textbf{actual,
physical experiments} that, as we show in the sequel, contribute
significantly to the overall success and efficiency of any closed-loop
autonomous experimental campaign. For example, in addition to
information-theoretic considerations typically considered in
decision-making policies, we ought to make other, more operational or
problem-specific ones when selecting experiments to run. In contrast to
hyperparameter optimization, where there is no additional cost incurred
when selecting vastly different parameters in succession, physical
experiments often do incur such a penalty, such as different times
required to increase versus decrease temperature of a furnace, or the cost of
selecting one type of catalyst material versus another, or the
effort and time needed to select one characterization method versus
another. Often, the time to run an experiment, wait for a system to
equilibrate, or wait for mechanical failure to occur must be decided 
as well, and many off-the-shelf models and methods do not elegantly
incorporate this. Much of our work focuses on modeling such aspects of
running physical experiments as rigorously as possible. In the next
section, we focus on three examples of such modeling.

\section{Higher features of materials experiments}
\label{sec:complex}

In developing more problem-fluent closed-loop autonomous materials
platforms, we have focused on better modeling problem-specific structure
and prior knowledge, and decision-making methods that better integrate
operational costs and rewards. Below, we list a few such methods and
extensions we have employed in the past.

\subsection{Incorporating physical models}
Perhaps one of the most striking departures we may take away from
generic, off-the-shelf closed-loop techniques such as black-box BO is
the ability to hypothesize or otherwise specify physics-based,
mechanistic insight relating the experimental actions with observed
experimental responses. By including physical models, we can impose a
more-specific structure on our predictions of response surfaces or other
unknown quantities, in contrast to the generic continuity or smoothness
assumptions that come equipped with purely statistical models such as
GPs. By integrating these models with Bayesian techniques, we can
additionally provide the qualification needed on such prediction to
reflect uncertainties arising due to model inaccuracies and noisy data.
Such an uncertainty quantification of predictions derived from a
physics-based model could subsequently inform decision-making policies
to select the next experiment. Such hybrid physical/statistical models
and decision-making that use them reside the box labeled ``Physics-informed
models" from the left in figure \ref{fig:scales}C.

One example of using physical models in the decision-making closed loop
is presented in \cite{chen2015optimal}. Here, we performed a simulation
study in which we considered the optimization of synthesis and
processing conditions to create a nanoemulsion. These conditions were
optimized against a measure of the stability of such an emulsion,
capturing two aspects. First, we desired the synthesized emulsion to
exhibit long-term stability at room temperature. Second, we required the
emulsion to be effectively destabilized at a high temperature. An
emulsion satisfying both criteria could be used to deliver payload
molecules contained in such an emulsion in a controlled, thermally
activated fashion.

In \cite{chen2015optimal}, we derived a kinetic model based on a system
of ordinary differential equations describing the destabilization of
such an emulsion, writing the objective experimental response function
in terms of controls, here the synthesis and processing conditions, and
a few effective physical parameters to the model, including energy
barriers and rate prefactors determining the kinetics of emulsion
instability \(f_M(\mathbf x; \bm{\theta})\). Under the assumption that
the model reflected the physical truth, up to the true value of the
effective physical parameters \(\bm{\theta}^\star\), we imposed Bayesian
beliefs on these parameters through a finite sample of \(L\) candidate
parameter values
\(\Theta = \left\{\bm{\theta}_1, ..., \bm{\theta}_L\right\}\). This
discretization of physical parameter space affords us simplicity in
approaching the Bayesian statistical calculations, effectively turning
the problem of assigning Bayesian beliefs into one of Bayesian
hypothesis testing
\cite{bernardo2002bayesian, schonbrodt2017sequential}, with each of the
\(L\) candidate parameter values severing as a hypothesis of what the
unknown, ground-truth parameter values are. At any point throughout the
campaign, we maintain the probabilities that each candidate is indeed
the ground-truth
\[p_i = \mathbb P\left[\bm{\theta}^\star = \bm{\theta}_i\right]\]
As new data is received, Bayes's law tells us how to update these
probabilities.

These probabilities along with the physical model of the response
\(f_M(\textbf x; \bm{\theta})\) also provide a sample of candidate,
mechanistic forward models \(f_i = f_M(\mathbf x; \bm{\theta}_i)\), each
qualified with the probabilities \(p_i\) that they represent the
ground-truth, unknown response
\(f^\star(\mathbf x) = f_m(\mathbf x; \bm\theta^\star)\). This sample
set of candidates, along with probabilities representing Bayesian
beliefs on such candidate, is a type of Bayesian Model Averaging (BMA)
\cite{hoeting1999bayesian}. With BMA, we can utilize physical models
within a Bayesian setting. With such BMA-based beliefs, we use the
Knowledge Gradient (KG) decision policy, which selects the experiment
\(\mathbf x\) --- here a set of synthesis conditions --- that maximize
the expectation
\begin{equation}
\nu_\text{KG}(\mathbf x) = \mathbb E\left[ \max_{\mathbf x^\prime} f^{n+1} (\mathbf x^\prime) - \max_{\mathbf x^\prime} f^n(\mathbf x) \right]
\label{eqn:kg}
\end{equation}
where \(f^n\) is the current mean-estimate of the response and
\(f^{n+1}\) is the future mean-estimate of the response if we were to
run experiment \(\mathbf x\) and the expectation is the weighted average
over the potential response outcomes of such an experiment. This
expectation is sometimes called an acquisition function, and different
such functions define other policies.

\begin{figure}
    \centering
    \includegraphics[width=0.95\textwidth]{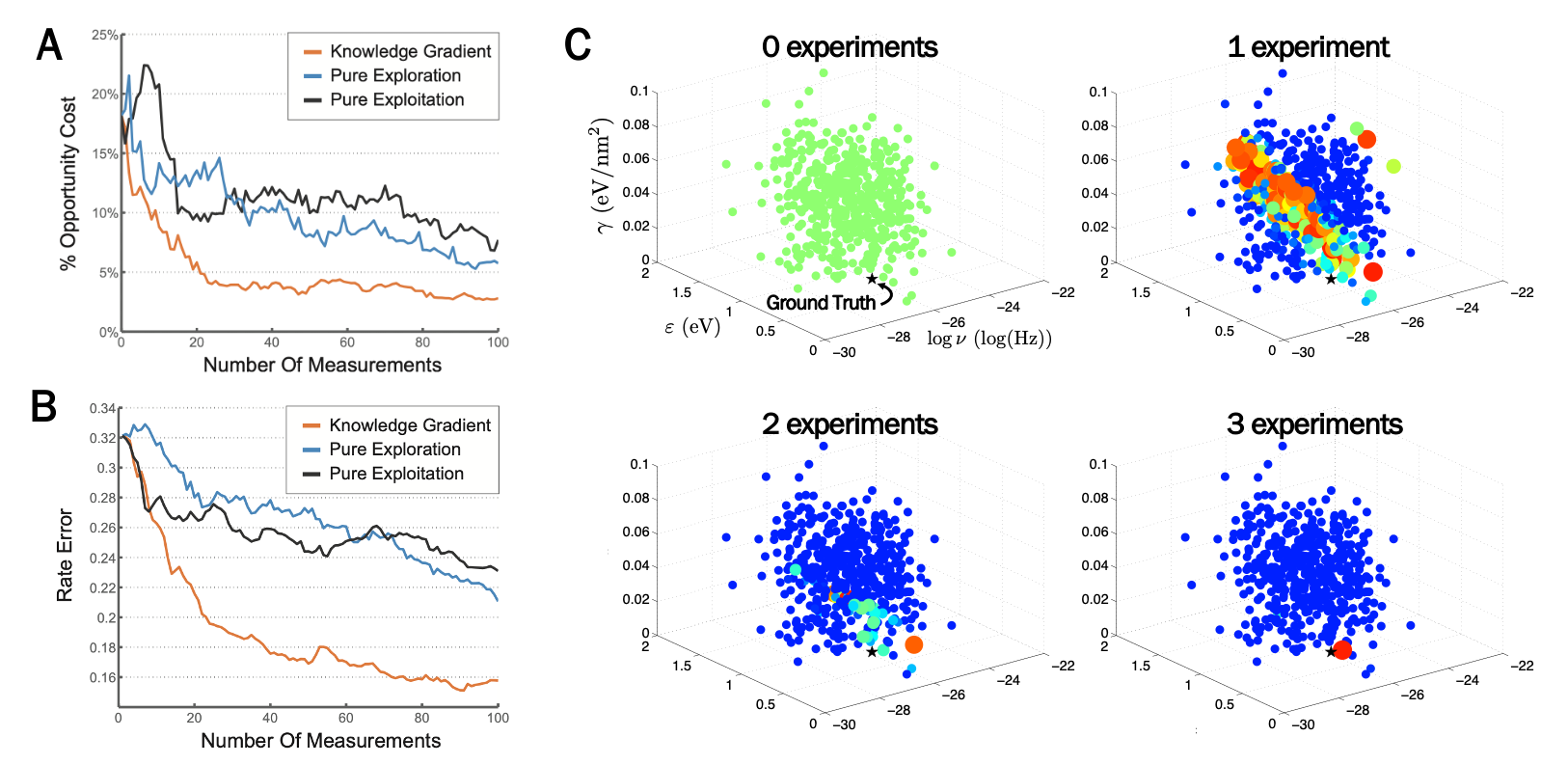}
    \caption{A) Percent opportunity cost and B) coalescence rate estimation
    error versus number of experiments for KG and baseline policies, from
    \cite{chen2015optimal}. C) Evolving Bayesian beliefs on physical
    parameters. Each dot is a sample of three physical parameters to a kinetic
    model: an interfacial energy $\gamma$, an activation energy $\varepsilon$
    and a reaction rate prefactor $\nu$. Larger, redder points correspond to
    higher posterior probabilities, while smaller, bluer points indicate smaller
    posterior probabilities.}
    \label{fig:chen}
\end{figure}

We run simulations to show how effective a closed-loop campaign that
uses both the BMA beliefs and the KG decision-policy is in finding the
synthesis conditions that optimize the response. This is plotted in
figure \ref{fig:chen}A, which plots a performance metric called percent
Opportunity Cost (OC, figure \ref{fig:closed_loop_design}B), calculated
and averaged over several simulations, where simulations sampled
different ground-truth values of parameters \(\bm{\theta}^\star\). 
OC reflects how much better the true optimal expected response value is
compared to the expected response value for the experiment believed to be
optimal, reflecting a penalty we pay for not picking the true optimal
experimental conditions. Lower OC indicates better performance in identifying
optimal or near-optimal synthesis conditions. We also compared the KG policy
against baseline policies. Compared to such policies, the KG policy was
significantly more effective in identifying the optimal set of synthesis
conditions over various ground truth scenarios.

Because of our use of parameterized physics models, the candidate
parameters \(\bm{\theta}_i\) along with the probabilities \(p_i\) allow
us to calculate weighted averaged estimates of the effective parameter
values \(\bm{\theta}^\star \approx \sum_{i=1}^L\bm{\theta}_i p_i.\) As
such, the evolving probabilities obtained after successive iterations of
the closed-loop provide increasingly refined insight regarding the
physical processes underpinning the observed experimental responses. For
example, in figure \ref{fig:chen}B, we plot the error of the mean
estimate of the droplet coalescence rate, which is a key kinetic process
contributing to emulsion stability. This error is a comparison between
the rate as determined by the averaged estimate
\(\sum_i \bm{\theta}_i p_i\) of the kinetic parameters and the true-rate
calculated from the ground-truth parameter values \(\bm \theta^\star\),
averaged over all simulations performed. We see that, in comparison to
the baseline policies, KG more efficiently learns the rate of this
process.

The use of physics-based models, when coupled with BMA and policies such
as KG, show the promise of deriving physical insight while
simultaneously optimizing material properties or experimental responses.
This insight becomes more effective if the set of candidates is selected
to represent a diverse set of models and mechanistic explanations of
what could potentially contribute to the observable response. How to
optimally generate and hypothesize such a set of models is an open topic
of research, and is a subject of some of our recent work. In
\cite{he2020optimal}, we specifically address two major points of
concern with the above BMA model. First is the assumption of a globally
applicable physical model \(f_M(\textbf x, \bm{\theta}).\) We explore
the mixture of physics-based models with statistical ones such as a GPs
through the use of local semi-parametric models, combine together
through a Gaussian kernel. In this way, we can consider fitting
reduced-order local models individually for smaller regions of
experiment space \(\mathcal X\). Such models may, for example, encompass
a smaller set of physical processes than a global model.

The second point addressed in \cite{he2020optimal} is selecting models
to include in the set \(\Theta\) of candidate models. One typically
encountered phenomenon when using BMA is the apparent dominance of one
candidate after a small number of observations
\cite{monteith2011turning}. This is illustrated in figure
\ref{fig:chen}C, which visualizes BMA beliefs on three-dimensional
physical parameter space of a simplified kinetic model derived from that
used in \cite{chen2015optimal}. Here the three physical parameters
describe an interfacial energy \(\gamma,\) an activation energy
\(\varepsilon\) , and a reaction rate prefactor \(\nu\). Each point in
the figure represents a single candidate parameter vector
\(\bm{\theta_i} = (\gamma_i, \varepsilon_i, \log{\nu_i})\) , while the
star represents the ground truth parameter. The sizes and colors
assigned to each point indicate the probabilities \(p_i\), with larger,
redder points representing larger probabilities, and smaller, bluer
points representing smaller ones. The figure shows the evolution of
beliefs after performing experiments selected using the KG policy. After
three data-points, one choice of parameters, hence one model dominates
--- the posterior probability associated with it is effectively 1. In
contrast, the other parameters have a corresponding posterior
probability of 0. While not exactly correct, the dominant parameter is
relatively close to the true parameter values. This represents a
coarse-grain refinement of our beliefs to identify a promising region in
parameter space. This identification allows us to resample a new set of
parameters effectively. Here we face several decisions regarding how to
perform such a resampling. For example, should we include models only in
the local neighborhood of the dominant model, or should we appreciate
the fact that this model is dominant in the face of a small number of
noisy data points? In \cite{he2020optimal}, we attempt to address such
questions and develop algorithms for coarse-grained and fine-grained
refinement of the set of candidate models in the face of such model
uncertainty.

In summary, within a Bayesian context, we can utilize knowledge about
physical, mechanistic models that predict responses we observe experimentally. 
Such a hybrid physical / Bayesian models can then be used with decision-making
policies to select experiments that balance between exploration and
exploitation. Due to our use of physics models, exploration amounts to more
than resolving statistical uncertainties, which is the case when using black-box
empirical models such as GPs, placing the technique one level removed from the
most empirical methods outlined in figure \ref{fig:scales}C.  Instead, we can
perform inference about the true mechanisms underpinning observed
responses.  In this way, we can use such a campaign to optimize synthesis
conditions while simultaneously learning about the physics of the material
system under study.

\subsection{Decision structure and meta-decisions in materials experiments}
Another form of problem-specific structure considers how decisions are
made and additional meta-decisions that impact an experimental campaign.
In \cite{wang2015nested}, we explore the problem of optimizing the
output current of an optoelectronic device created by attaching
nanoparticles to a photosensitive substrate. The design variables are
the size, shape, and composition of the nanoparticles, in addition to
the density of the nanoparticles on the substrate. While finding the
optimal design variables that maximize output current could be addressed
using problem-agnostic BO methods and algorithms, such a solution
disregards the fact that the nanoparticles take time to synthesize.
Also, nanoparticle density can be varied in a high-throughput manner by
deposing NPs with a density gradient on the substrate. As such, it is
costly to vary the size, shape, and composition of the NPs, but several
NP densities can be tested simultaneously and in parallel. Such a
problem would occupy the space between the boxes labeled ``Batch / multi-stage
decisions" and ``Operational, contextual, or exogenous variables" in figure
\ref{fig:scales}C due to the consideration of experimental structure, and the
operational constraints such structure poses.

\begin{figure}
\centering
    \includegraphics[width=0.65\textwidth]{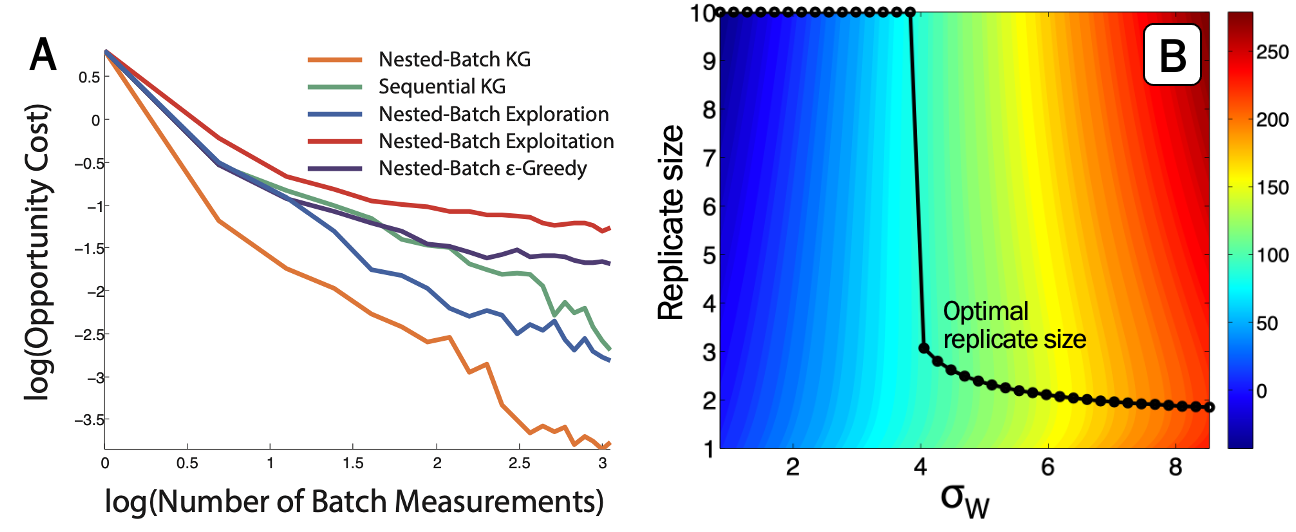}
    \caption{A) Opportunity cost versus number of experiments for nested-batch
    mode policies, taken from \cite{wang2015nested}.   B) Total cost (color
    value) versus single-period noise standard deviation and number of
    replicates, and the optimal replicate size (black circles and lines) for a
    given noise level, taken from \cite{reyes2015quantifying}.}
    \label{fig:wang}
\end{figure}

This imposes a nested and batched structure to the decisions. First, we
must decide on a specific nanoparticle type by specifying shape, size,
and composition. Once synthesized, we may consider a batch of NP
densities as a second, nested decision. In \cite{wang2015nested}, we
showed how to augment current BO policies, such as the KG policy, to
incorporate this nested-batch decision-making structure. This is done
through a simulation-optimization procedure that used Monte Carlo
simulations to sample and select design parameters over which to compute
the expectation needed to calculate the KG acquisition function value
defined in equation \eqref{eqn:kg}. Figure \ref{fig:wang}A shows
simulation results comparing nested-batch KG policy with other baseline
policies, showing the augmented nested-batch KG policy outperforming the KG
policy that does not properly model the nested-batch nature of the decisions
being made, and nested-batched versions of other baseline policies.

As illustrated above, the cost of running any particular experiment is a
reality of physical experimentation that is not properly modeled using
normal closed-loop design methods such as BO. In
\cite{reyes2015quantifying} we explore this further through an analysis
of meta-decisions made in an experiment and the costs associated with
such decisions. We consider archetypal experimental design
meta-decisions such as the number of experiment replicates performed,
such as performing experiments in duplicate or triplicate. By averaging
the responses of such an ensemble of near-identical experiments,
scientists hope to reduce the impact of noise. In fact, if we assume an
experimental response is perturbed through an additive Gaussian noise as
in equation \eqref{eqn:noise}, then assuming this noise has variance
\(\sigma^2_W,\) the averaged response over \(n\) independent replicates of the
experiment should have a reduced variance of \(\sigma_W^2/n\).  However, this
replication comes at an experimental cost, which depends on the
parallelizability available to run such an \(n\)-fold replicated study. If done
sequentially, the effectively reduced variance is obtained with a factor of
\(n\) additional experiments. In contrast, perfectly parallel experiments do
not incur any additional experimental cost compared to running a single
experiment. This cost of running \(n\) experiments can be captured through the
formula
\[C_\alpha(n) = \alpha + (1-\alpha)\cdot n\]
where \(\alpha\) is a number between 0 and 1 that serves as a measure of
parallelizability. In \cite{reyes2015quantifying}, we consider the
impact of a campaign's total cost with respect to this and other
hyper-parameters such as measurement noise and campaign stopping
criteria. For example, in \ref{fig:wang}B, we plot the total cost of a
campaign needed to find reduce the opportunity cost to within 7\%,
assuming a parallelizability coefficient \(\alpha = 0.6\) using the KG
policy. This total cost of a campaign is plotted with respect to the
single-experiment noise \(\sigma_W\) and the replicate size \(n\).

The black dots and connecting lines indicate the optimal replicate size
for any given amount of noise. This is the replicate size that minimizes
the total cost of the campaign. Counter-intuitively, these simulations
show that for this problem, as the single-experiment noise becomes
larger, it is advantageous to select a smaller number of identical
experiments per each iteration of the closed-loop campaign. One way to
understand this result is to acknowledge that the KG policy often
performs relatively better in high-noise scenarios than other policies.
Reducing the replicate size to 1 simply in such scenarios allows the KG
policy to be more agile in adaptively selecting \(n\) different
experiments rather than \(n\) identical experiments for this amount of
parallelization.

\subsection{State of the campaign and operational considerations}
The last example of problem-specific structure considers the case of
operational considerations of running an experimental campaign. Such
considerations allow us to consider and optimally select operational
variables that impact some notion of reward or costs of an experiment
and consider the campaign's termination criteria based on such costs.
One common example of such considerations and termination criteria is
the time to perform a particular experiment and the overall time-budget
of a campaign. Another example is considering the probability of failure
in running an experiment. Very often, such costs and termination
constraints depend on maintaining a more complicated nature of the
state, apart from the simple state-of-belief that we have used thus
far.  For example, if we are concerned about a fixed time-budget for running an
autonomous experimental campaign, then we must consider how long we
have spent in the current campaign whenever we decide on the next action.
Problems concerning this and other operational considerations would
fall in the top-right box of figure \ref{fig:scales}C.

Here, we employ the generic framework of Reinforcement Learning (RL) and
Markov Decision Processes (MDPs), which is not a large departure from
the one presented in section \ref{sec:review}. The main differences here
is the notion of the state of a campaign, a criterion to determine when
a campaign is terminal and should be stopped, and the formulation of the
decision-policy based on some measure \(R(S, \mathbf x, y)\) of the
rewards obtained or costs --- sometimes called regret --- incurred after
taking a particular action \(\mathbf x\) from a particular state \(S\)
and observing a specific outcome \(y\). Given a sequence of states,
\(\left\{S_i\right\}\) actions taken by such states according to some
decision-making policy \(\left\{\mathbf x_i = \pi(S_i)\right\}\) and the
resultant experimental outcomes from such actions
\(\left\{y_i\right\}\), we can define the total cumulative reward
\[\sum_{i} R(S_i, \mathbf x_i, y_i),\]
One of the goals in reinforcement learning is to select the
\(\mathbf x_i\), i.e., learn the policy that maximizes this total
cumulative reward on average. Key to this is the concept of the value of
a state \(V^\pi(S)\), which we may think of as the total additional
rewards we can expect to obtain starting from this state \(S\) and
proceeding with our experimental campaign using policy \(\pi\) until we
arrive at some terminal state. Thus, like KG and EI, the value of a
state is the expectation of some future quantity that measures the
quality of a particular state. However, unlike KG and EI, the value
\(V^\pi\) considers modeling not only the outcome of a single future
experiment but the outcome of an entire experimental campaign. It is
computationally infeasible to calculate such values exactly, and many RL
techniques such as value-function iteration, policy-function iteration
\cite{sutton2018reinforcement, powell2007approximate} and Q-learning
\cite{watkins1992q}, are essentially centered around approximating this
or other related functions. The result is an approximation of
\textbf{the} optimal policy, i.e. the one geared toward maximizing
cumulative reward or minimizing cumulative regret.

Under this framework, rather than identifying and testing different
policies, we instead define different reward structures. This provides a
more flexible method for including constraints and operational
considerations. For example, if we wished to optimize a material
response subject to the constraint that we cannot surpass a particular
time-budget for a campaign, we can specify the time-budget as a
termination criterion and specify a single reward obtained at the end of
a campaign equal to the maximum/minimum of the predicted response
overall potential experimental inputs
\[R(S_i, \mathbf x_i, y_i) = \begin{cases} 
\max \mu_{S_i}(\mathbf x) & \text{ if $S_i$ is terminal. } \\ 
0 & \text{ otherwise. } \end{cases}\]
where \(\mu_{S_i}(\mathbf x)\) is the mean estimate of the response
given we are in state \(S_i\).

\begin{figure}
\centering
    \includegraphics[width=0.9\textwidth]{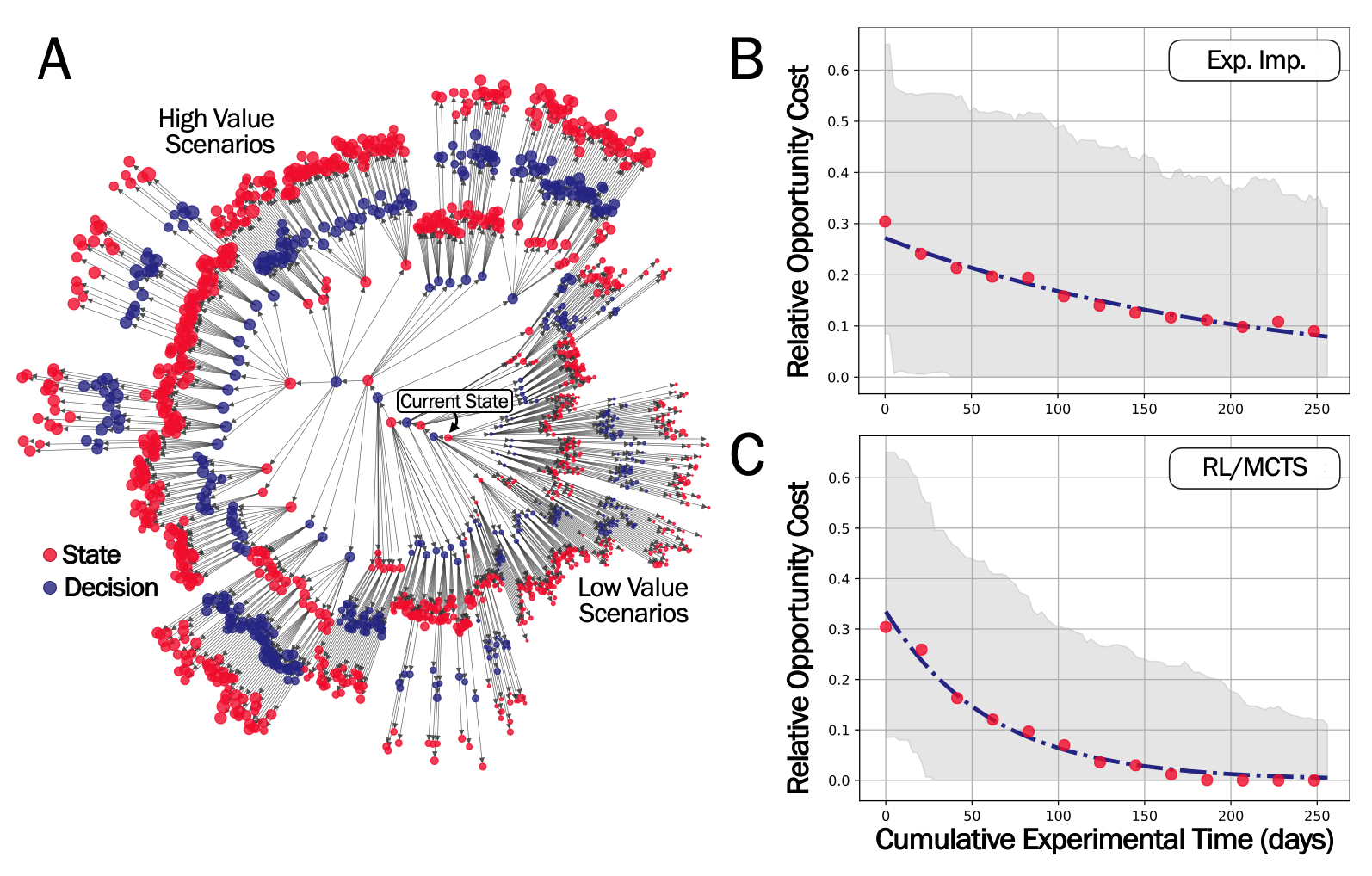}
    \caption{A) A scenario tree sampled using Monte Carlo Tree Search,
    enumerating high-value future scenarios of an experimental campaign. Red
    vertices are correspond to simulated future states, and blue vertices are
    simulated future decisions. Vertex size indicate the value assigned to the
    state or action. B) and C) Opportunity cost versus cumulative experimental
    time using a BO policy and a MCTS/RL policy, respectively. Red dots indicate
    median OC values obtained over several simulations, and dotted blue line
    is an exponential decay fit. Gray regions indicate the envelope between
    25th and 75th percentiles of OC values.}
    \label{fig:rl}
\end{figure}

One method we employ to calculate state values and make decisions within
this RL framework is Monte Carlo Tree Search (MCTS). MCTS is a method
for sampling experimental campaign scenarios that potentially result in
high-value states \cite{browne2012survey, coulom2006efficient}. This is
done through sequentially building up a tree of scenarios that enumerate
various simulated experimental decisions, outcomes, and future states.
Having such a scenario tree, we can approximate the utility of pursuing
any particular branch, which results in a decision-making policy: select
the action leading down the branch with the highest expected cumulative
rewards. An example scenario tree is depicted in figure \ref{fig:rl}A.
Here the blue vertices denote instances of a simulated decision being
made during some simulated experimental campaign. Red vertices reflect
the state obtained after making such a decision and obtaining a
simulated outcome. The size of the nodes indicates the value associated
with a particular state or action. We observe a discrepancy of values
for different scenarios, which informs what actions to pursue from the
actual current state of the campaign, which is highlighted in the
figure.

Such a framework allows us to include more problem-specific operational
considerations in our decision-making. For example, we may consider the
following scenario, motivated by experiments performed at the
Relativistic Heavy Ion Collider at Brookhaven National Laboratory
\cite{bnl1988relativistic}. One such experiment measures the
distribution of net-baryon number fluctuations produced as a function of
the collider's beam energy, as part of the Beam Energy Scan program,
which is attempting to search for a conjectured critical point of the
phase diagram of quantum chromodynamics \cite{luo2017search}. In
particular, identifying minima and other features of moments of this
distribution, such as the kurtosis of the distribution \(\kappa(E)\),
seen as an experimental response as a function of beam energy \(E\),
help scientists in this critical point search
\cite{PhysRevD.96.074510, aggarwal2010higher, thader2016higher}.

Identifying the optimal energy \(E\) that minimizes the net-baryon
kurtosis response function \(\kappa^\star(E)\) can be approached using
vanilla BO methods. However, such methods do not appropriately
appreciate an operational choice of experiment time. That is, scientists
at RHIC, in addition to energy, must specify the time. In principle, the
ground-truth net-baryon distribution is not impacted by experiment time.
However, the observed estimate of this net-baryon distribution does
depend on experimental time in addition to energy. This discrepancy
between the ground-truth kurtosis \(\kappa^\star(E)\) and the observed
kurtosis can be modeled as heteroscedastic noise, so that we can write
the observed kurtosis response \(\hat \kappa(E, t)\) obtained when
running an experiment with beam energy \(E\) and experiment time \(t\)
as
\[\hat{\kappa}(E,t) = \kappa^\star(E) + W(E, t)\]
where \(W(E,t) \sim \mathcal N(0, \sigma^2_W(E, t))\) is Gaussian noise
with variance \(\sigma^2_W(E,t)\) that depends on \(E\) and \(t\). One
model assumes that the noise level is inversely proportional to \(E^3\)
and \(t\). That is, noise decreases at higher energies and longer
experiment times. Thus there is a trade-off between experimental
accuracy and experimental time. Proper, problem-fluent policies that
appreciate this trade-off will balance selecting a large number of
potentially inaccurate experiments versus a small number of accurate
experiments. However, vanilla BO policies such as EI do not make such a
consideration.

Figure \ref{fig:rl}B shows the results of several hundred simulated
closed-loop BO campaigns run to find optimal energies of sampled
ground-truth kurtosis response functions using GP beliefs for
\(\kappa^\star\) and an EI policy. The figure plots the median, 25th,
and 75th percentiles of relative OC as a function of experiment campaign
time, rather than the number of experiments. We see that after a
simulated campaign time of 256 days, the EI policy can find, on average,
energies producing kurtosis response that within 10\% of the optimal,
minimal kurtosis response. In contrast, by using MCTS-based RL policy,
cognizant of the time budget of 256 days, and the above noise model, we
obtain the simulation results depicted in figure \ref{fig:rl}C. Here we
see that this RL-policy obtains a similar optimality level in around 80
days, which reflects a campaign speed-up factor of more than 3.

The key difference between BO-based policies and RL-based ones is that
BO-policies make information-theoretic considerations designed to
minimize the number of iterations around the closed-loop. With RL-based
policies, we can add more complex rewards and regrets that reflect
additional measures of experimental costs and impose campaign
termination constraints that reflect budgets on those costs. Simulations
of entire campaigns can use such constraints to more effectively and
holistically plan the next experiment with the remaining campaign budget
in mind. While the above example of experiments at the RHIC may seem
exotic, the general tradeoff between experiment time and accuracy is
common throughout experimental science. This trade-off exists when
deciding how long to run an experiment to obtain a reasonable,
extrapolated estimation of some material property. This could arise when
trying to extrapolate a quantity of a material or chemical system at an
equilibrated state, such as predicting equilibrium chemical
concentrations through the determination of reaction rates. This also
arises when extrapolating for out-of-equilibrium or long timescale
events and properties, like failure times of a mechanical component.
Using RL-based policies, we can better include this trade-off when
planning autonomous materials campaigns.

\section{Conclusion}
We have outlined a few examples of how we have approached closed-loop
autonomous materials design through a common framework that involves
Bayesian beliefs about the system being studied and decision-making
policies that select optimal experimental actions to run. We have shown
examples of how black-box models and methods can accelerate materials
research but have indicted three samples of how such techniques are
deficient. By integrating problem-specific structure, we showed a few
examples of how we rectify such deficiencies.

As ML models and methods become more commonplace in materials science
research, we have the opportunity to enrich traditionally data-driven,
problem agnostic techniques with problem-specific structure. Doing so in
the context of ML for materials could result in smaller data
requirements, more robust models, and the ability to draw interpretable
scientific insight. This opportunity is well-pronounced in the new area
of autonomous materials development, where combining data-driven
statistical models with physics and operational considerations result in
a more meaningful, scientifically richer autonomous materials
exploration in which robot scientists have the agency to pursue and make
higher-order decisions and formulate more complex knowledge of the
material system being studied.  

The idea of combining physical, statistical, and operational models to form a
such continuum of methods appropriate at different scales of knowledge and
decision-making capabilities (figure \ref{fig:scales}C) is analogous
to the continuum of computational materials methods that straddle length and
time scales (figure \ref{fig:scales}A). As computational material
scientists, we understand that no one method is globally appropriate.  By
understanding the benefits, risks, and trade-offs between models at different
scales, we can tackle materials modeling problems more effectively.  So too, we
believe, can we effectively apply ML and statistical models and methods through
an appreciation of their strengths and weaknesses, supplementing the latter
with more problem-fluent approaches.

\section{Acknowledgments}
This review features work resulting from helpful conversations and
collaborations with several groups. Specifically, we acknowledge Si
Chen, Yan Li, Yingfei Wang, Xinyu He and Warren Powell from Princeton
University, Frank Alexander, Kevin Yager, Jaimie Dunlop from Brookhaven
National Laboratory, and Benji Maruyama, Rahul Rao, Jennifer Carpena,
Ahmad Islam and Pavel Nikolaev from the Air Force Research Laboratory,
Keith Brown and Aldair Gongora at Boston University, and Milad Abolhasani
and Robert Epps at North Carolina State University.

\section{Data Availability}
The raw data required to reproduce these findings are available to download
from 
\begin{center}
https://www.csms.io/data
\end{center}
The processed data required to reproduce these findings are available to
download from 
\begin{center}
https://www.csms.io/data
\end{center}

\bibliography{mybib}
\bibliographystyle{ieeetr}

\end{document}